\documentstyle [12pt]{article}
\begin{document}

\hfill {WM-95-111}

\hfill {December, 1995}

\vskip 1in   \baselineskip 18pt
{
\Large

   \bigskip
   \centerline{IMPLICATIONS OF A HIGGS DISCOVERY AT LEP} }
 \vskip .8in

\centerline{P.Q. Hung${}^1$ and Marc Sher${}^2$}
\bigskip
\centerline {${}^1$\it Physics Department, University of Virginia,
Charlottesville, VA 22901, USA}
\centerline {${}^2$\it Physics Department, College of William and
Mary, Williamsburg, VA 23187, USA}

\vskip 1cm

{\narrower\narrower  If the Higgs boson has a mass below 130 GeV,
then the standard model vacuum is unstable; if it has a mass below 90 GeV
(i.e. within reach of LEP within the next two years), then the instability will
occur at a scale between 800 GeV and 10 TeV.  We show that precise
determinations
of the Higgs and top quark masses as well as more detailed effective potential
calculations will enable one to pin down the location
of the instability to an accuracy
of about 25 percent.  It is often said that ``the standard mo
del must break down''
or ``new physics must enter'' by that scale.
However, by considering a toy model
for the new physics, we show that the lightest new particle (or resonance)
could
have a mass as much as an order of magnitude greater than the location of the
instability, and still restabilize the vacuum.}

\newpage

\def\beq{\begin{equation}}
\def\eeq{\end{equation}}
\def\lm{\lambda}
\def\Lm{\Lambda}

\section{Introduction}
The large value of the top quark mass has intensified interest in
Higgs mass bounds arising from the requirement of vacuum
stability.  Since the contribution of the top quark Yukawa
coupling to the beta function of the scalar self-coupling, $\lm$,
is negative, a large top quark mass will drive $\lm$ to a negative
value (thus destabilizing the standard model vacuum) at some
scale, generally denoted as $\Lm$.  The only way to avoid this
instability is to require that the Higgs mass be sufficiently large
(thus the initial value of $\lm$ is large) or to assume that the
standard model breaks down before the scale $\Lm$ is
reached\cite{sherrep}.

If one assumes that the standard model is valid up to the
unification or Planck scale (the difference between the two does
not appreciably affect the bounds), then a lower bound on the Higgs
mass can be obtained by requiring that the standard model vacuum be
the only stable minimum up to that scale.  Many papers in recent
years\cite{sherlet,ceq,altarelli} have gradually refined this lower
bound; the most recent are the works of Casas,
Espinosa and Quiros\cite{ceq}[CEQ] and of Altarelli and
Isidori\cite{altarelli}[AI],
 who show that the requirement of vacuum stability up
to the Planck scale gives, for a top quark mass of 175 GeV, a lower bound of
130
GeV on the Higgs mass.  If the Higgs mass is lighter than this bound, then the
standard model must break down at a lower scale; the farther below the
bound, then the lower this scale.  In fact, as emphasized by
AI and CEQ, if the Higgs has a mass just
above its current experimental limit, then the standard model must
break down at a scale of roughly a TeV.  Since the standard model
is {\it defined} by the assumption that there is no new physics
until a scale of several TeV (at least), one concludes that the
discovery of a Higgs boson at LEPII could, depending on the
precise top quark mass,  rule out the standard model!!

In this Letter, we will examine this question in more detail.
First, we will discuss the scale, $\Lm$, at which the Higgs
potential turns negative, thus destabilizing the
standard model vacuum.  We will consider Higgs masses within reach
of LEPII.  Then, the uncertainties and difficulties associated with
determining this scale precisely, given the Higgs and top quark
masses, will be discussed.  The standard statement is that ``the
standard model must break down before the scale $\Lm$" or that
``new physics must operate before the scale $\Lm$". We will examine
the meaning of this statement by considering a toy model in which a
scalar boson of mass $M$ is added to the standard model, and we
will show that it is {\it not} necessary that this mass be less
than $\Lm$,  that it could even have a mass as high as $5-10$ times
larger and still prevent the vacuum instability.  Thus, even if one
were to conclude that $\Lm$ is only a TeV, this would not
necessarily imply that a particle or resonance must be at or below
this scale.

\section{The Higgs potential and location of the instability}

For top quark and Higgs masses of interest, the Higgs potential
has its usual electroweak minimum at $246$ GeV, and then at some
larger scale, $\Lm$, turns around (sharply) and becomes negative,
destabilizing the electroweak vacuum\footnote{We will define $\Lm$
to be the point at which the potential drops below the value of
the electroweak minimum, however the drop is so rapid that this
does not appreciably differ from the point at which it turns
around.}.  In order to calculate the potential as accurately as
possible, one must sum all leading and next-to-leading
logarithms.  This is done\cite{kast} by improving the one-loop
effective potential by two-loop renormalization-group equations.
The most recent and detailed calculations of the bounds are those of Altarelli,
et al.\cite{altarelli} and Casas, et al.\cite{ceq}.  The reader is referred to
those papers for details, we will simply present their results here.

The procedure is straightforward.  One begins with values of the scalar
self-coupling, $\lm$, and the top quark Yukawa coupling,  evaluated
at some scale
(usually $m_Z$).  One then integrates these using the two-loop renormalization
group equations. The running couplings are then inserted into the one-loop
Higgs potential, which is then examined to see if it goes negative, and if so,
at what scale.  Finally, the Yukawa coupling and $\lm$ must be converted into
pole masses for the physical Higgs boson and top quark.  Many issues involving
the choice of renormalization scale and the renormalization procedure must be
considered\cite{ceq,ford}.

The most important results of the papers of AI and CEQ was the bound on the
Higgs mass assuming that the standard model is valid up to the Planck scale.
They obtained (using a value of the strong coupling given by
$\alpha_s(m_Z)=0.124$):
\beq m_H\ >\ 130.5 + 2.1(m_t - 174) \eeq
for AI (all masses are in GeV) and
\beq m_H\ > \ 128 + 1.92(m_t - 174) \eeq
for CEQ.  These results are in agreement to well within the stated $3-5$ GeV
errors of the two calculations.

Suppose one looks at the values of $m_H$ and $m_t$
which give an instability at $\Lm=1$ TeV.  AI obtain\footnote{The results of
CEQ differ significantly, by as much as 15 GeV.
The reason appears to be related
to the work of Willey and Bochkarev\cite{willey1}.  They noted that the
contribution of the finite $\overline{MS}$ electroweak tadpoles to the relation
between the top quark pole mass and the mass defined in terms of the
$\overline{MS}$ Yukawa coupling is much larger than the well-known QCD
correction.  A similar contribution exists in relating the Higgs pole mass to
the potential.  Willey\cite{willey2} has pointed out that this contribution
cancels in the Higgs-top mass ratio.
In the paper of CEQ, it was included in the Higgs
mass relation, but not in the top mass relation.  By including the term in the
top mass relation in the CEQ work, Willey\cite{willey3} has found that the
discrepancy becomes much smaller (and the agreement for large $\Lm$ persists).}
\beq m_H = 71 + .9(m_t-174)\eeq

We have reproduced this result  and have
generalized it to different $\Lm$.  We find that the
factor of $.9(m_t-174)$ is unchanged, and the factor
of $71$ GeV changes to $77$ GeV for $\Lm=1.5$ TeV, $81$ GeV for $\Lm=2$ TeV,
and $89$ GeV for $\Lm=4$ TeV.  Thus, knowing the
experimental values of the  Higgs mass and top quark masses to an accuracy of 1
GeV each, which may be possible at an NLC, will enable one to determine $\Lm$
to
roughly 25 percent accuracy.

Before discussing the significance of knowing a particular value of $\Lm$, it
is important to discuss the uncertainty in this formula.  As discussed very
clearly by CEQ, the choice of the scale-dependence of the renormalization scale
introduces uncertainties of roughly $3-5$ GeV in the Higgs mass.  In addition,
one must be very careful in determining the condition for the instability.
For example, AI took the instability to occur when
$\lm$ became negative, whereas CEQ took the instability to occur when
$\tilde{\lm}$, given by (ignoring, for illustrative purposes, the electroweak
gauge couplings)
\beq
\tilde{\lm}=\lm-{1\over 32\pi^2}\bigg[6h^4_t\left(\ln{h^2_t\over
2}-1\right)\bigg]\eeq goes negative,
which minimizes uncertainties due to higher orders.  This
can make a small difference in $\Lm$, as CEQ show, which is irrelevant for
large
$\Lm$ but can be very important for smaller $\Lm$.

Finally, there is another, potentially serious, uncertainty.  In running the
scalar self-coupling from $m_Z$ to the location of the instability, one
includes
top quark contributions at all scales, and thus the beta function is
negative at $m_Z$ and drives the scalar self-coupling towards
negative values immediately.  However, suppose one were to argue
that the top quark loop contributions to the beta  function should
not enter until $2m_t$  is reached, as is usually the case for
the QCD beta function.  Then, $\lm$ would not change much between
$m_Z$ and $2m_t$, increasing the location of the instability
by (very roughly) a factor of $2m_t/m_Z\sim 4$.  Of course, the
$\overline{MS}$ renormalization scheme is mass-independent, and
thus the contribution should be included at all scales, but this
does indicate that another renormalization scheme which is more
sensitive to threshold effects could give significantly different
results.  This would imply that uncalculated higher order
contributions could become important if threshold effects are
included.

In principle, all of these issues can be dealt with (and certainly will be if
the Higgs boson is discovered at LEP).
In that case, the results of Eq. 3 will be
accurate to within a couple of GeV.  If the
Higgs boson is discovered next year at
LEP, then the standard model vacuum will be known to be unstable at a scale of
somewhere between
$0.8$ and
$10$ TeV.  The biggest uncertainty in pinning down this number is the top
quark mass.  Once it is known to an accuracy of around $5$ GeV, then the
biggest uncertainty will be in the above calculations.  When these
uncertainties are removed, then the location of the instability, $\Lm$, will
be known to roughly a factor of 2.  Finally, as the experimental values of
the Higgs and top quark masses are narrowed down to  1 GeV each, the location
of the instability will eventually be determined to roughly $25\%$ accuracy.

We now consider the following question.
Let us suppose that this happens, and one
concludes that the instability occurs at, say, 1000-1400 GeV.  What does this
imply for new physics?  Must a new particle or resonance occur with a mass
below or near this scale?

\section{Model of New Physics}

In order to examine the effects of new physics, a simplified
version of the standard model Higgs potential will be considered
in which the renormalization scale dependence of the parameters is ignored.
The resulting potential can then be written as
\beq
V=-{1\over 2}m^2\phi^2+{1\over 4}\lm\phi^4-{1\over
4}B\phi^4\left(\ln{\phi^2\over\mu}-C\right)\eeq
where
\beq
B={1\over 16.\pi^2}\left(3h_t^4-3g^4/8-3(g^2+g^{\prime 2})^2/16.
\right)\eeq
This simplified potential has all of the qualitative features of
the full renormalization-group improved potential, and the results
obtained from its use will not be substantially changed by using
the full potential.  Differentiating the potential, one can
replace $m^2$ and $\lm$ by the Higgs mass $m_h$ and the electroweak
minimum, $\sigma$:
\beq
V=-{1\over 4}m^2_h\phi^2-{1\over 2}B\sigma^2\phi^2+{3\over
8}B\phi^4+{1\over 8\sigma^2}m^2_h\phi^4-{1\over
4}B\phi^4\ln{\phi^2\over \sigma^2}\eeq
Plugging in a top quark mass of 190 GeV\footnote{The pole mass is
chosen to be 190 GeV, so that the Yukawa coupling corresponds to a
mass which is 5-6 \% smaller.}, and a Higgs mass of 70 GeV, one
can see that the potential turns around and becomes negative at a
scale, $\Lm$, of $1250$ GeV.  This would seem to imply that ``new
physics" must enter by that scale.

In order to model the ``new physics", we will add to the model a
scalar field, with bare mass $M$, which couples to the standard
model Higgs field with coupling $\delta$ (so that the mass-squared
of the scalar is $M^2+\delta\sigma^2$).  In addition, the
multiplicity of the scalar field will be $N$.  This is fairly
general.  We know that additional fermionic degrees of freedom
will further destabilize the vacuum, so that only bosonic degrees
of freedom need to be considered.  These degrees of freedom must
couple to the Higgs field (to have any effect on the potential),
and one would expect a number of such fields (if they are vector
fields, of course, the multiplicity of each would be 3).

What are reasonable values for $N$ and $\delta$?  The value of
$N$ will be taken to be anywhere between $1$ and $100$.  Such
large values of $N$ are not implausible.  In the minimal
supersymmetric model, for example, the multiplicity of scalar
quarks is $N=72$ ($6$ for flavor, $3$ for color, $4$ for
particle/antiparticle and left/right); in left-right models, the
multiplicity of the new gauge bosons and Higgs bosons is $N\sim 25$.
$\delta$ will be taken to be between $0.1$ and $10$.  In the next
section, the unitarity bound on $N$ and $\delta$ will be found and
we will only assume that the values must be lower than that bound.
It is plausible that the value of $\delta$ would be close to the
unitarity bound, if the effective new physics is strongly coupled.

The effects of the scalar on the Higgs potential is to add a term
\beq
{N\over 64\pi^2}(M^2+\delta\phi^2)^2\left(\ln{M^2+\delta\phi^2
\over \mu^2}-C\right)\eeq
to the potential.  Differentiating the potential, one
can replace $m^2$ and $\lm$ with $m_h$ and $\sigma$, yielding
\begin{eqnarray}
V &=& -{1\over 4}m^2_h\phi^2-{1\over 2}B_1\sigma^2\phi^2
+{3\over 8}B_2\phi^4+{1\over 8\sigma^2}m^2_h\phi^4
-{1\over 4}B\phi^4\ln{\phi^2\over \sigma^2}\nonumber\\
&+& {N\over 64\pi^2}(M^2+\delta\phi^2)^2\ln{M^2+\delta\phi^2
\over M^2+\delta\sigma^2}\end{eqnarray}
where
\begin{eqnarray}
B_1&=&B-{N\over
32\pi^2}(2\delta^2-\delta{M^2\over\sigma^2})\nonumber\\
B_2&=&B-{N\over 16\pi^2}\delta^2.\end{eqnarray}

This potential can be plotted, and examined for various values of
$N$, $\delta$ and $M$ to see if the instability remains.  Some
features are easy to see.  Consider the limit in which $M=0$, so
the additional scalars are very light.  In this case, the
coefficient of the $\phi^4\ln\ \phi^2$ term is
$N\delta^2/64\pi^2-B/4$.  Thus, if $N\delta^2$ is too small, this
coefficient will be negative and the instability will remain.  Thus, a
lower bound on $N\delta^2$, in order to remove the instability, is
\beq
N\delta^2\ >\ 3h^4_t-3g^4/8-3(g^2+g^{\prime 2})^2/16.\eeq
It is also interesting to consider the limit in which
$M\rightarrow\infty$.  In this case, the logarithm can be expanded
and one can see that the effects of the extra term vanishes
completely, as expected from the decoupling theorem.  In this
case, the scalar field will not restabilize the potential,
regardless of the values of $N$ and $\delta$.

Thus, for any given values of $N$ and $\delta$ (above the
critical value), there will be some critical value of the scalar
mass; if $M$ is below this value, the potential will be
restabilized; if $M$ is above this value, it will not be.
This is illustrated in Fig. 1.  We have chosen $N=60$ and
$\delta=1$, and have plotted the potential for various values
of $M$.  In this case, $M$ can be as large as $5.3$ TeV, and
still restabilize the potential.  Note that this value for $M$ is
four times the value of $\Lm$, the point by which ``new physics
must enter".

The critical value of the scalar mass is shown as a function of
$N$ and $\delta$ in Fig. 2.  Note that the $M=0$ line corresponds
to the critical value of $N\delta^2$ shown above.  We see that for
the largest values of $N$ and $\delta$, the scalar mass could be
as large as 100 times $\Lm$, and still restabilize the potential!
Of course, one would question the validity of perturbation theory
for such values, and we now turn to the question of the unitarity
bounds on $N$ and $\delta$.

\section{Unitarity Bounds}

There are two types of unitarity bounds that we can consider; a
bound on $\delta$ from tree-level unitarity and a bound on
$N\delta^2$ from one-loop unitarity.

 The first is the bound on
$\delta$ arising from the requirement of tree-level unitarity.  If
we call the scalar $S$ and the Higgs boson $H$, then one will
obtain a bound on
$\delta$ by requiring that the real part of each $H\ S\rightarrow\
H\ S$ partial wave scattering amplitude be less than
$1/2$\cite{unitarity}.  This calculation can be easily done, and
in the limit that the quartic $S^4$ coefficient is small, we find
the bound $\delta<4\pi$.  This is a fairly weak bound, which is
weaker than the bound obtained from the following bound on
$N\delta^2$.

A bound on $N\delta^2$ will arise by considering $H\ H\
\rightarrow\ H\ H$ at one-loop in which a loop with $S$ bosons
is in the diagram--this will be proportional to
$N\delta^2/8\pi^2$; so one might expect a bound on $N\delta^2$
somewhat less than $8\pi^2$.  One-loop unitarity is a more difficult
problem, and the bound will always depend on
$\sqrt{s}$ of the scattering process\cite{durand}. We will estimate
the bound in two very different ways.

The first, and simplest, method is to note that the beta function
for $\lm$ can be read off from the potential of Eq. (9), and
clearly has a term proportional to $N\delta^2$.  Thus, we can
 integrate $\lm$ from the electroweak scale (or the Z
mass--the choice doesn't significantly affect the result) up to the
scale given by $M$, and simply require that $\lm$ not exceed its
unitarity limit by that scale.  Since we are using one-loop beta
functions here, we only require that $\lm$ not exceed its tree
level unitarity bound, given by  $\lm\sim 4$ (this value
corresponds to a Higgs mass of $700$ GeV, which is the
Lee-Quigg-Thacker bound\cite{unitarity}).  When we do so, we
find the bound given by the solid line of Fig. 2, which
corresponds to $N\delta^2$ varying between 30 and 60.

The second method is to compute the scattering  amplitude for $HH
\rightarrow HH$ at one loop. Since we are interested in $m_h \leq 90$
GeV, the Higgs self-coupling
$\lambda$ is small ($ \leq 0.067$) and  the one-loop contributions to
the above process coming from $H$ and the Goldstone  bosons $w,z$ can be
neglected. The dominant contribution comes from $S$. In the limit that
the quartic $S^4$ coefficient is small compared with $\delta$, we can
ignore the one-loop contribution to $HS \rightarrow HS$.

The one-loop contribution of $S$ to $HH \rightarrow HH$ can be
straightforwardly computed. The renormalization consists of two parts.
One comes from the one-loop self energy of $H$ due to $S$ where a factor
of $N \delta^2$ is present. (We are again ignoring the contributions due
to
$H$ and $w,z$ which are proportional to $\lambda^2$.) This contributes
to the wave function renormalization constant for $H$ and to the
renormalization of $\lambda$.  The other comes from the bubble diagram
for $HH \rightarrow HH$  involving $S$. The final physical scattering
amplitude is, of course, finite. In the limit $\sqrt{s}\ >>\  M,m_h$,
the real and imaginary parts of the S-wave partial wave amplitude,
$a_0$, are  given by
\begin{eqnarray} {\rm Re}\  a_0 &=& -(\frac{3}{8 \pi}) \lambda_s +
\frac{N \delta^2}{64 \pi^3} (1+3 m_h^2 I_s^{\prime} (m_h^2)),\nonumber\\
{\rm Im}\ a_0 &=& \frac{N \delta^2}{128 \pi^2},
\end{eqnarray} where
\begin{equation}
\lambda_s = \lambda + \frac{N \delta^2}{8 \pi^2} (\ln\frac{\sqrt{s}}{M}
-1 +
\frac{1}{2} I_s(m_h^2)),
\end{equation} with $I_s(p^2) =\int_0^1 dx\ \ln(1+4 x (1-x)/\beta)$ and
$\beta = 4 M^2/p^2$ ($I_s^{\prime}$ is the derivative of $I_s$ with
respect to $p^2$). Numerically, $I_s(m_h^2)$ and $m_h^2 I_s^{\prime}
(m_h^2)$ are small.

By including the (tree level) real S-wave amplitude ($\delta/ 8 \pi$)
for $HS \rightarrow HS$ and diagonalizing the real part of the $2 \times
2$ matrix, we can plot the Argand diagram for the largest eigenvalue
(see Durand, et al.\cite{durand} for a detailed discussion). The upper limits
on
$N\delta^2$ are found by looking at the point where the amplitude deviates
significantly from the unitarity circle.   We find the following results which
depend on
$\sqrt{s}$:
$N
\delta^2 <30$ (
$\sqrt{s}/M \approx 100$); $N \delta^2 < 60$ ($\sqrt{s}/M \approx 40$);
$N \delta^2 < 100$ ($\sqrt{s}/M \approx 20$). This corresponds
respectively to $\lambda_s = 1.42,\ 2.07, \ 2.58$. This approach gives
results which are basically consistent (within a factor of two in
$N\delta^2$) with the ones obtained by ``running" $\lambda$ as discussed
above.

\section{Conclusions}

{}From Fig. 2, we see that a single scalar boson with a coupling
$\delta\sim 6$ to the standard model Higgs (which is at, but not
above, the unitarity bound) can have a mass as high as 10 TeV,
and still succeed in eliminating the instability which would
occur at 1250 GeV.  If there were a strongly interacting sector,
one might expect just such a coupling.

Thus, should LEP discover a Higgs boson in the near future, one
will conclude that the standard model must ``break down" at some
calculable scale, $\Lm$, which could be between 1 and 10 TeV
(depending on the top quark mass).  In this letter, we have shown
that this does not necessarily mean that a new particle(s) or
resonance(s) must exist at this scale, but that the new states
could be close to a factor of 10 higher in mass.  There is no
guarantee that an accelerator which reaches the scale $\Lm$ will
find any direct evidence of new physics.

We are very grateful to Old Dominion University for its
hospitality while this work was carried out and thank Ray Willey for
many useful discussions.  This work supported by the National Science
Foundation, grant PHY-9306141, and by the Department of Energy,
grant DE-A505-89ER40518.

 \def\prd#1#2#3{{\it Phys. ~Rev. ~}{\bf D#1} (19#2) #3}
\def\plb#1#2#3{{\it Phys. ~Lett. ~}{\bf B#1} (19#2) #3}
\def\npb#1#2#3{{\it Nucl. ~Phys. ~}{\bf B#1} (19#2) #3}
\def\prl#1#2#3{{\it Phys. ~Rev. ~Lett. ~}{\bf #1} (19#2) #3}

\bibliographystyle{unsrt}

\newpage
\parindent=0pt
\section*{Figures}
\begin{enumerate}

\item
The Higgs potential is plotted for three different values
of $M$, with $N=60$ and
$\delta=1$.  Between the origin and $\phi=500$ GeV, the three curves are
essentially identical, and look like the conventional Higgs potential.  For
very
large $M$, the scalar field decouples
and the potential develops an instability at
$1250$ GeV.  As $M$ decreases to $5.5$ TeV, the instability point moves outward
and then disappears for $M= 5.3$ TeV.

\item
For various values of $N$ and $\delta$, the largest value of $M$ which will
eliminate the instability (which occurs at $1250$ GeV in the absence of the
additional scalar field).  The shaded region
covers the values of $N$ and $\delta$
which violate the unitarity bound, as discussed in the text.
\end{enumerate}

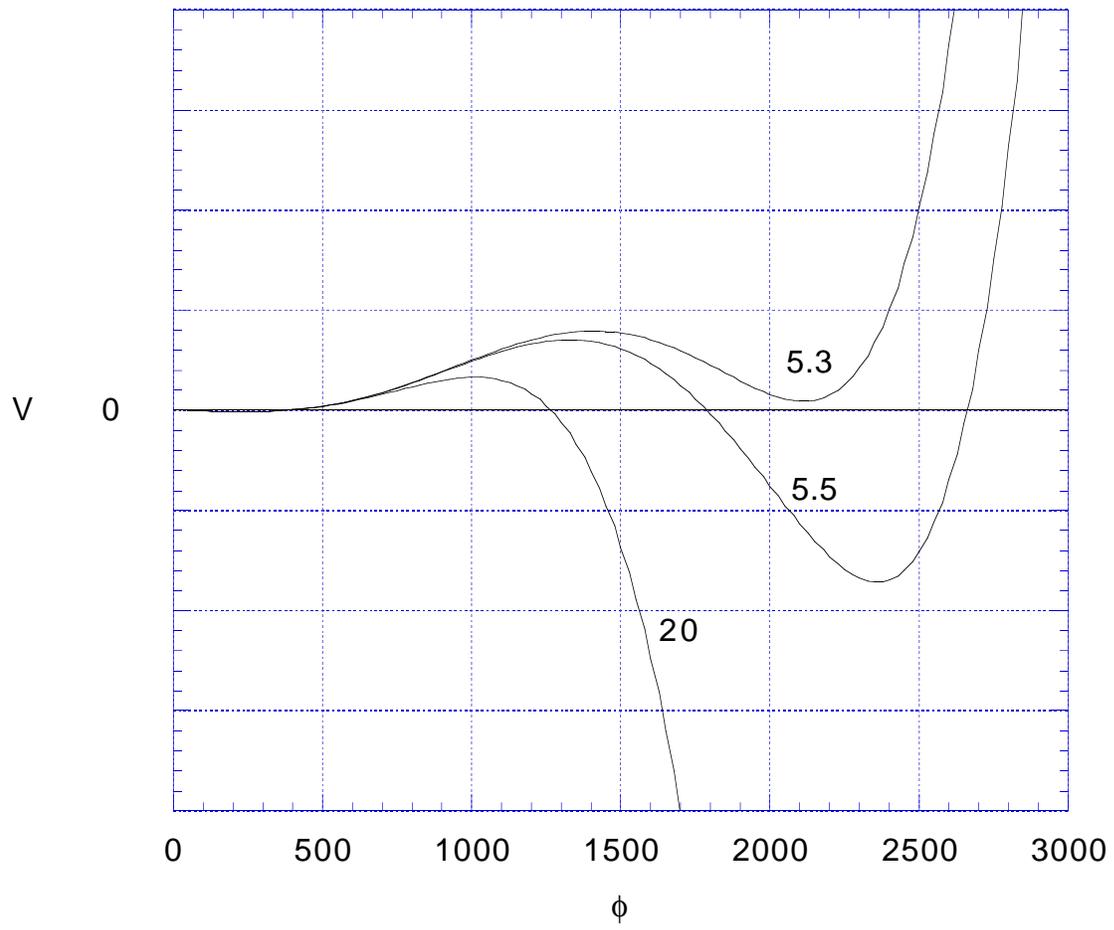
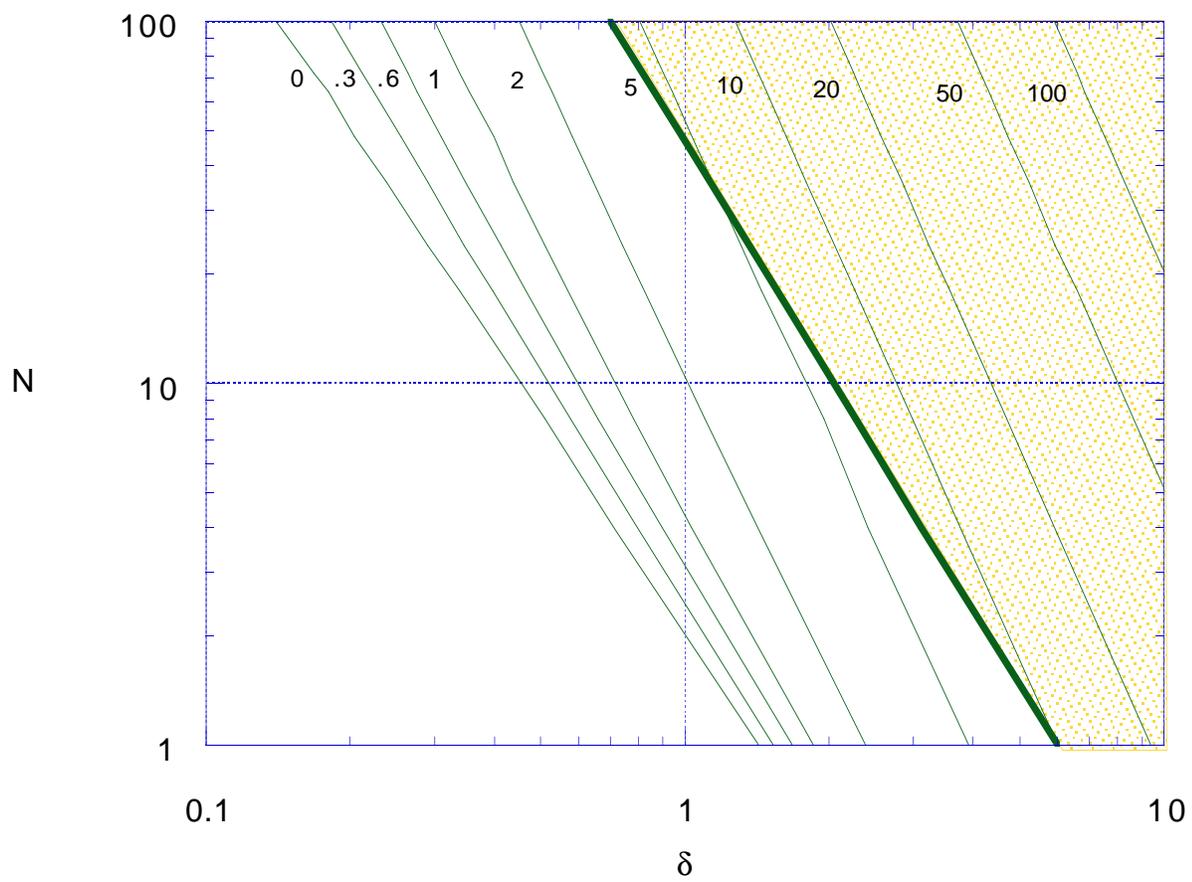
\end{document}